\documentclass[12pt]{article}
\usepackage{graphicx}
\usepackage{amsfonts}
\usepackage{amssymb}
\newcommand{\nin}{\noindent}
\newcommand{\be}{\begin{equation}}
\newcommand{\ee}{\end{equation}}
\newcommand{\bea}{\begin{eqnarray}}
\newcommand{\eea}{\end{eqnarray}}

\newcommand{\nn}{\nonumber\\}

\begin{document}

\begin{titlepage}

\nin KCL-PH-TH/2010-24

\begin{centering}
\vspace{1cm}
{\Large {\bf  Spherically symmetric solutions in Covariant  Horava-Lifshitz Gravity}}\\

\vspace{1cm}

{\bf Jean Alexandre $^{1,\dagger}$},  {\bf Pavlos Pasipoularides $^{2,\#}$}

\vspace{.2in}

$^1$ King's College London, Department of Physics,\\
London WC2R 2LS, UK

$^2$ Department of Physics, National Technical University of Athens, \\
Zografou Campus GR 157 73, Athens, Greece \\

\vspace{3mm}

\end{centering}
\vspace{0.5cm}

\begin{abstract}
We study the most general case of spherically symmetric vacuum solutions in the framework of the Covariant
Horava Lifshitz Gravity, for an action that includes all possible higher order terms in curvature which are compatible with power-counting normalizability requirement. We find that solutions
can be separated into two main classes: {\it (i)} solutions with nonzero radial shift function, and
{\it (ii)} solutions with zero radial shift function. In the case {\it (ii)}, spherically symmetric solutions are consistent
with observations if we adopt the view of Horava and Melby-Tomson \cite{Horava:2010zj}, according to which the auxiliary field $A$ can be considered as a part of an effective general relativistic metric, which is valid only in the IR limit. On the other hand, in the case {\it (i)}, consistency with observations implies that the field $A$ should be independent of the spacetime geometry, as the Newtonian potential arises from the nonzero radial shift function. Also, our aim in this paper is to discuss and compare these two alternative but different assumptions for the auxiliary field $A$.
\end{abstract}

\vspace{1.2cm}
\begin{flushleft}
$^\dagger$ jean.alexandre@kcl.ac.uk \\
$^\#$ paul@central.ntua.gr \\
\end{flushleft}
\end{titlepage}

\section{Introduction}

A recent higher order space derivative model for Gravity was formulated by Horava~\cite{Horava:2008ih}. This model is power-counting
renormalizable and serves as an ultraviolet (UV) completion of General Relativity (GR). This scenario is based on an anisotropy
between space and time coordinates, which is expressed via the
scalings $t\rightarrow b^z t$ and  $x\rightarrow b x$, where $z$
is a dynamical critical exponent. It is worth noting that Horava-Lifshitz (HL) Gravity has stimulated an extended research
on Cosmology and black hole solutions, see for example \cite{Kiritsis:2009sh}-\cite{Iorio:2009qx}, and
we note that quantum field theory models in flat anisotropic space-time were also developed,
see for example \cite{Horava:2008jf}-\cite{Orlando:2009az} and references therein.

In HL Gravity, the
four-dimensional diffeomorphism  invariance of GR
is sacrificed in order to achieve power-counting
renormalizability. Although HL Gravity violates local
Lorentz invariance in the UV, GR is expected to be recovered in the infrared (IR) limit. This implies a very
special renormalization group flow for the couplings of the model, in particular
it is expected that the coupling $\lambda$ in the extrinsic curvature term of the action has the behavior $\lambda \rightarrow 1$, i.e.
that it flows towards its GR value.
But there is no theoretical study supporting this specific
behavior. In addition, there are several other potential inconsistencies in
HL Gravity which have been discussed (see for example
\cite{Charmousis:2009tc}-\cite{Pons:2010ke}
and references therein). More specifically, the breaking of 4D diffeomorphism invariance
introduces an additional scalar mode which may lead to strong
coupling problems or instabilities, and in this way prevents HL Gravity from fully reproducing GR in the IR limit.

In Ref.\cite{Horava:2010zj} a new Covariant HL Gravity is formulated by Horava and Melby-Thompson (HM),
which includes two additional nondynamical fields $A$ and $\nu$, together with a new $U(1)$ symmetry.
In this model the $U(1)$ symmetry eliminates the extra scalar mode curing the strong coupling problems in the IR limit.
Note, that in what follows we adopt the HM action of Ref. \cite{Horava:2010zj}, in which the parameter $\lambda$ is set equal to one ($\lambda=1$). However, the $U(1)$ symmetry can not force the value of the parameter $\lambda$ to be equal to 1, as an action, with the extended gauge symmetry and $\lambda\neq 1$, has been formulated in \cite{daSilva:2010bm} . Also, HM theory reproduces many features of GR at long distances as it is shown initially in Ref.\cite{Horava:2010zj}.

HL Gravity can be separated into two versions which are known as projectable
and non-projectable. In the projectable version the lapse function
$N$ (see Eq. (\ref{ADM}) below) depends only on the time coordinate, while in the
non-projectable version $N$ is a function of both space and time coordinates.
The Covariant HL Gravity considers the projectable case.

In this paper we study the most general case of spherically symmetric vacuum solutions in the framework of Covariant HL Gravity,
for an action
\footnote{For the construction of the HL action,
the so called "detailed balance principle" has been proposed \cite{Horava:2008ih}.
The main advantage of this approach is the restriction of the
large number of arbitrary couplings that
appear in the action of the model. However,
a more general way for constructing the action
would be to include all possible operators which are compatible
with the renormalizability \cite{Kiritsis:2009sh,Sotiriou:2009bx}; this implies
that all operators with dimension less or equal to six are allowed.}
which includes all possible terms allowed by renormalizability requirement.
We find that spherically symmetric solutions can be separated into two main classes:
{\it (i)} solutions with nonzero radial shift function, and
{\it (ii)} solutions with zero radial shift function.

We would like to note that Covariant HL Gravity, as it is formulated by Horava and Melby-Tomson (HM) in \cite{Horava:2010zj} , incorporates an additional assumption for
the field $A$. In particular, the field $A$ is assumed as a part of an effective general relativistic metric in the IR limit,
via the replacement $N\rightarrow N-A/c^2$. Spherically symmetric solutions, in the case {\it (ii)}, are consistent
with observations if we adopt HM approach for $A$. On the other hand, in the case {\it (i)}, we see that $A$ should be
independent of the lapse function in order to achieve consistency with observation. We would like to warn the reader,
that these two alternative but different views for the field $A$ are examined both in this paper, in the framework of
spherically symmetric vacuum solutions of cases {\it (i)} and {\it (ii)}.

The paper is organized as follows. In Sec.\ref{sec2} we summarize features of the Covariant HL Gravity.
We consider then in Sec.\ref{sps} the most general ansatz for spherically symmetric solutions, including
a nonzero radial shift function, and we derive the equations of motion and the corresponding constraints. In Sec.\ref{as} we present the
solutions for the two situations {\it (i)} and {\it (ii)}, and finally Sec.\ref{con} contains our conclusions.
In Appendix A are presented some details on the geometrical interpretation of the auxiliary field $A$.
In Appendix B, we present the minimal substitution approach of Ref. \cite{daSilva:2010bm} which describes how mater couple with the auxiliary fields $A$ and $\nu$.

\section{Covariant Horava-Lifshitz Gravity } \label{sec2}

In this section we introduce the notation for the Covariant HL
Gravity in the case of three spatial dimensions ($d=3$), and we
discuss the additional $U(1)$ symmetry of the model, as well as the role of the auxiliary nondynamical fields
$A$ and $\nu$ which are included in the action.

\subsection{The action}

This model, as the original HL Gravity, is characterized by an anisotropy between space and
time dimensions
\begin{equation}
[t]=-z, \quad [x]=-1~,
\end{equation}
where $z$ is an integer dynamical exponent.The action of the model is structured by a set of
five fields: $N(t)$, $N_{i}(x,t)$, $g_{ij}(x,t)$, $A(x,t)$ and
$\nu(x,t)$  $(i=1,2,3)$. Note that $N(t)$, $N_{i}(x,t)$, $g_{ij}(x,t)$ are the standard fields that appear
in the Arnowitt, Deser and Misner (ADM) form of the space-time metric
\begin{equation} \label{ADM}
ds^2=-  c^2 N^2 dt^2+g_{ij} \left(dx^i+N^i dt\right)\left(dx^j+N^j
dt\right)~,
\end{equation}
where $c$ is the velocity of light, with dimension $[c]=z-1$, and spatial
components $dx^i/dt$ $(i=1,2,3)$. In addition, $N$ and $N_i$ are
the "lapse" and "shift" functions which are used in general
relativity in order to split the space-time dimensions, and
$g_{ij}$ is the spatial metric of signature (+,+,+). Note that
here we are interested for the projectable version of the model which
implies that the lapse function $N(t)$ depends only on the time parameter. For the
dimensions of "lapse" and "shift" functions we obtain
\begin{equation}
[N]=0, ~~[N_i]=z-1~.
\end{equation}
The auxiliary fields $A(x,t)$ (potential) and $\nu(x,t)$ (prepotential
\footnote{The terminology "potential" for $A(x,t)$ and "prepotential" for $\nu(x,t)$ has been introduced in the original work
of \cite{Horava:2010zj} } ) are
nondynamical fields which have to satisfy constraint equations. As we will see subsequently
the existence of these fields is necessary in order to
achieve invariance of the action under the extended Gauge symmetry: $U(1)\times Diff({\cal M},{\cal F})$. In addition, the dimensions of
these fields are
\be
[A]=2z-2,~~[\nu]=z-2~.
\ee
The full action of the model is formulated as
\bea\label{action}
S&=&\frac{2}{\kappa^2}\int dt d^3x\sqrt{g}\Big\lbrace N\left[ K_{ij}K^{ij}-K^2-V+\nu\Theta^{ij}(2K_{ij}+\nabla_i\nabla_j\nu)\right]\nn
&&~~~~~~~~~~~~~~~~~~~~~~~~~~~~~~~~~~~~~~~~~~~~~~~~-A(R-2\Omega)\Big\rbrace ,
\eea
in which $d$ is the spatial dimension ($D=d+1=4$), $\kappa^2$ is an overall
coupling constant with dimension $[\kappa^2]=z-d$, and the extrinsic curvature is
\be
K_{ij}=\frac{1}{2 N} \left\{\dot{g}_{ij}-\nabla_i N_j-\nabla_j N_i\right\},~~ i,j=1,2,3
\ee
where the symbol $\Theta^{ij}$ is defined as
\be
\Theta^{ij}=R^{ij}-\frac{1}{2}R g^{ij}+\Omega g^{ij}~.
\ee
Note that this choice for $z=3$ is an immediate consequence of
power counting renormalizability request. In particular, the coupling $\kappa^2$ in the above action
has dimension $[\kappa^2]=z-3$, hence, if $z=3$ the HL Gravity model is renormalizable, for $z>3$ is super-renormalizable and
for $0<z<3$ is non-renormalizable.

For the construction of the potential term $V$ we will
not follow the standard detailed balance principle, but we will
use the more general approach \cite{Kiritsis:2009sh, Sotiriou:2009bx}, according to
which the potential term is constructed by including all possible
renormalizable operators (relevant and marginal) \footnote{We have ignored terms which
violate parity, see also \cite{Sotiriou:2009bx}.}, that have dimension smaller
than or equal to six, hence we write
\be \label{pot1}
V= V_{IR}+V_{R^2}+V_{R^3}+V_{\Delta R^2}
\ee
where
\bea \label{pot2}
&& V_{IR}=-c^2 (R-2\Lambda)\\
&&V_{R^2}=-\alpha_1 R^2-\alpha_2 R^{ij}R_{ij} \nn
&&V_{R^3}=-\beta_1 R^3-\beta_2 R R^{ij} R_{ij}-\beta_3 R_i^j R_j^k R^i_k\nn
&&V_{\Delta R^2}=-\beta_4 R\nabla^{2}R-\beta_5 \nabla_i R_{jk} \nabla^i R^{jk}\nonumber
\eea
The dimensions of the various terms in the Lagrangian are
\be
[R]=2,~ [{R^2}]=4, ~[{R^3}]=[{\Delta R^2}]=6,
\ee
where the symbol $\Delta$ is defined as $\Delta=g_{ij}\nabla^i\nabla^j$ ($i=1,2,3$).
In addition, we have used the notation $R$, $R_{ij}$ and $R_{ijkl}$
for the Ricci scalar,  the Ricci and the Riemann tensors
($i,j=1,2,3$), which correspond to the spatial 3D metric
$g_{ij}$. Note that the term $ R^{ijkl}R_{ijkl}$ does not appear in $V_{R^2}$, as the Weyl tensor in three
dimensions automatically vanishes. In addition, $c^2$, the couplings $\alpha_i$ ($i=1,2$), and
$\beta_j$ ($j=1,\cdots,5$), have dimensions
\be
[c^2]=4~~,~~[\alpha_i]=2~~,~~
[\beta_j]=0.
\ee
Finally, we would like to note that the potential terms, of Eqs. (\ref{pot1}) and (\ref{pot2}) above,  
has been considered previously for covariant HL gravity, in the case of cosmology, by the authors of Ref. \cite{Wang:2010wi}.

\subsection{The extended $U(1)\times Diff({\cal M},{\cal F})$ Gauge symmetry}

The main motivation for considering models with an anisotropy between space and
time dimensions type is the construction of a power-counting renormalizable
Gravity model. However,  in order to achieve normalizability,
and simultaneously keep the time derivatives up to second order,
we have to sacrifice the standard 4D diffeomorphism invariance of
General Relativity, which is now restricted to
\be \label{diffeo}
\delta t=f(t),~~ \delta x^{i}=\xi(t,x^{j})
 \ee
which is a foliation preserving diffeomorphism, $ Diff({\cal M},{\cal F})$, where
${\cal M}$ is the spacetime manifold, provided with a preferred foliation structure ${\cal F}$.
In particular the fields $N,~N_i,~g_{ij}$ transforms as:
\begin{eqnarray}
\delta g_{ij}&=&\partial_{i}\xi^{k}g_{ik}+\partial_{j}\xi^{k}g_{ik}+\xi^{k}\partial_{k}g_{ij}+f\dot{g}_{ij} \\
\delta N_{i}&=&\partial_{i}\xi^{j}N_{j}+\xi^{j}\partial_{j}N_{i}+\xi^{j}g_{ij}+\dot{f}N_i+f \dot{N}_{i} \nonumber \\
\delta N&=&\xi^{j}\partial_{j}N+\dot{f}N+f \dot{N} \nonumber
\end{eqnarray}
However, the action of Eq. (\ref{action}) has an additional symmetry, in particular
it remains invariant under a $U(1)$ Gauge symmetry, according to which the fields of the model transform as
\begin{eqnarray}\label{gauge}
\delta_\alpha N&=&0\nonumber \\ \delta_\alpha g_{ij}&=&0 \nonumber\\
\delta_\alpha N_{i}(x,t)&=&N \nabla_{i}\alpha \nonumber \\ \delta_\alpha A(x,t)&=&\dot{\alpha}-N^{i}\nabla_{i}\alpha\nn
\delta_\alpha\nu&=&\alpha
\end{eqnarray}
where $\alpha$ is an arbitrary spacetime function. Accordingly, the full symmetry of the action of Eq. (\ref{action})
is the extended Gauge symmetry:  $U(1)\times Diff({\cal M},{\cal F})$.

The extrinsic curvature term in the action of Eq. (\ref{action}), if we assume only $Diff({\cal M},{\cal F})$ symmetry,
could include an additional running coupling constant $\lambda$ appearing as:
\be \label{curvature1}
 {\cal L}_{K}= K^{ij}K_{ij}-\lambda K^2,
\ee
In order to achieve agreement with General Relativity, we expect that RG flow in the IR leads the coupling
$\lambda$ to unity, but the exact mechanism for this remains unknown. There was a hope, in Ref. \cite{Horava:2010zj}, that the Covariant model
requires $\lambda=1$ due to the $U(1)$ symmetry, hence the above mentioned problem, for the flow of $\lambda$
in the IR, does not exist. On the other hand, in Ref. \cite{daSilva:2010bm},  it was shown that the action of Eq. (\ref{action}) can be
written in an invariant form under $U(1)$ symmetry, for an arbitrary value of $\lambda$, which implies that
$U(1)$ symmetry can not fix the coupling $\lambda$ (see Appendix B in the present paper).

Note that according to HM approach an infinitesimal $U(1)$ transformation on the fields $A,\nu,N,N^i$, in
the nonrelativistic limit, is equivalent to an infinitesimal diffeomorphism involving the time coordinate. As a consequence,
the symmetry U$(1)\times$ Diff(${\cal M},{\cal F},)$ can be seen approximately as a Diff({\cal M},4)
symmetry of Standard General Relativity, see Ref. \cite{Horava:2010zj} and the discussion in our Appendix A.

\subsection{Infrared limit}

Although the additional $U(1)$ symmetry eliminates the extra degree of freedom, the IR limit of the HM theory,
obtained after neglecting higher order in spatial curvature terms, does not coincide with General Relativity. Indeed,
the action is then
\bea \label{IR}
S&=& \frac{1}{16\pi G} \int dx^0 d^3 x \sqrt{g}
\Big\{N\left[K_{ij}K^{ij}- K^2 + R +\nu\Theta^{ij}(2K_{ij}+\nabla_i\nabla_j\nu)\right]\nn
&&~~~~~~~~~~~~~~~~~~~~~~~~~~~~~~~~~~~~~~~~~~~~-A(R-2\Omega)\Big\}
\eea
where the time-like coordinate $x_{0}$ is defined as $x_{0}=c t$, and the fields (in the above action) are
rescaled according to Sec.5 in Ref. \cite{Horava:2010zj}.
However, as discussed in \cite{Horava:2010zj}, Covariant HL Gravity reproduces many features of general
relativity for long distances.

\section{Spherically symmetric solutions in Covariant HL Gravity} \label{sps}

\subsection{The metric}

The starting point is the action \cite{Horava:2010zj}, describing Gravity with anisotropic scaling:
\bea\label{action1}
S&=&\frac{2}{\kappa^2}\int dt d^3x\sqrt{g}\Big\lbrace N\left[ K_{ij}K^{ij}-K^2-V+\nu\Theta^{ij}(2K_{ij}
+\nabla_i\nabla_j\nu)\right]\nn
&&~~~~~~~~~~~~~~~~~~~~~~~~~~~~~~~~~~~~~~~~-A(R-2\Omega)\Big\rbrace ,
\eea
We consider the most general static spherically symmetric metric, of the form:
\be\label{metric}
ds^2=-c^2 N^2 dt^2+\frac{1}{f(r)}\left(dr+n(r)dt\right)^2+r^2(\sin^2\theta d\theta^2+d\phi^2),
\ee
where $n(r)=N^{r}(r)$ is the radial component of shift functions, and $N_r=n(r)/f(r)$ since $g_{rr}=1/f(r)$.

\subsection{Constraints}

The auxiliary fields $\nu$ and $A$ lead to the following constraints
\begin{itemize}
\item The variation of $S$ with respect to $A$ gives $R-2\Omega=0$, or equivalently
\be
R=-\frac{2}{r^2}\left(r f'+f-1\right)=2 \Omega~,
\ee
which imposes that the function $f(r)$ in the metric (\ref{metric}) is
\be\label{solf}
f(r)=1-\frac{\Omega}{3}r^2-\frac{2 B}{r}~,
\ee
where $B$ is a constant of integration. As we will see, in the following sections, this
constant of integration $B$ is not interpreted necessary as the mass of the spherically compact
object. Moreover, in what follows, we will set $\Omega=0$, as we are looking for asymptotically flat
solutions.
\item The variation of $S$ with respect to $\nu$ gives
\be \label{const2}
\Theta^{ij}\nabla_i\nabla_j\nu+\Theta^{ij}K_{ij}=0,
\ee
In what follows we will assume the Gauge fixing $\nu=0$, then the above constraint gives $\Theta^{ij}K_{ij}=0$, which is satisfied for
spherically symmetric solutions as we will see in the following sections.
\end{itemize}

\subsection{Equations of motion}

We therefore start with the action
\be\label{S1}
S\sim\int dtd^3x\sqrt{g}\Big\lbrace N\left( T-V\right)-A R\Big\rbrace, ~~T=K_{ij}K^{ij}-K^2  ,
\ee
in which we have considered the Gauge fixing $\nu=0$, and we have set $\Omega=0$.
The Lagrangian which corresponds to the action (\ref{S1}), has the form
\begin{eqnarray}
{\cal L}&=&{\cal L}_{T}+{\cal L}_{V}-\frac{r^2}{\sqrt{f(r)}} A R,\\
{\cal L}_{T}&=&\frac{N(t)r^2}{\sqrt{f(r)}} T(r,n,n',f,f'),\nn
{\cal L}_{V}&=&-\frac{N(t) r^2}{\sqrt{f(r)}} V(r,f,f',f'',f'''),\nonumber
\end{eqnarray}
where
\begin{eqnarray}
T&=&-\frac{8}{N^2(t) r^2}(n^2f^2+2 r n f^2 n'+r n^2 f f') \nonumber \\
-\frac{r^2}{\sqrt{f(r)}}A R&=&\frac{2}{\sqrt{f(r)}} A(r) (r f'+f-1).
\end{eqnarray}
The potential term of the Lagrangian, for zero cosmological term $(\Lambda=0)$ (as we are interested for solutions which are
asymptotically flat), is taken as
\bea\label{potential}
V=&-&c^2 R-\alpha_1R^2-\alpha_2R^{ij}R_{ij}-\beta_1R^3-\beta_2RR^{ij}R_{ij}\\
&-&\beta_3R^i_{~j}R^j_{~k}R^k_{~i}-\beta_4 R\nabla^{2}R- \beta_5\nabla_i R_{jk} \nabla^i R^{jk},\nonumber
\eea
in which
\bea
R&=&-\frac{2}{r^2}\left(r f'-1+f\right) \nn
R^{ij}R_{ij}&=&\frac{1}{2r^4}\left[ 3(rf')^2+4rf'(f-1)+4(f-1)^2\right] \nn
R^i_{~j}R^j_{~k}R^k_{~i}&=&-\frac{1}{4r^6}\left[5(r f')^3+6(r f')^2( f-1)+12 r f' ( f-1)^2+8 (f-1)^3\right] \nn
R\nabla^{2}R&=&\frac{2(r f'-1+f)}{r^6}\times\nn
              && \left[2r^3 f''' f+r^3 f'' f'+2 r^2 f''f+ 2 (1-3 f) r f'-4f(1-f) \right] \nn
\nabla_i R_{jk} \nabla^i R^{jk}&=&\frac{f}{2 r^6} [3(r^2f'')^2-2 r^3 f' f''+8 r^2 f'' (1-f)+ \nn
&&+5 (r f')^2+16 (1-f)r f'+24(1-f)^2] \nonumber.
\eea
The Euler equation for $n$ gives
\be
\frac{d}{dr}\left(\frac{\partial {\cal L}_{T} }{\partial n' }\right)=\frac{\partial {\cal L}_{T}}{\partial n},
\ee
from which we obtain
\be
f'(r) n(r)=0,
\ee
such that necessarily either $n(r)=0$ or $f(r)$ is constant. The Euler equation for $f$ is
\be
\sum_{n=0}^3(-1)^{n}\frac{d^n}{dr^n}\left( \frac{\partial{\cal L}}{\partial f^{(n)}}\right) =0,
\ee
where $f^{(n)}=d^nf/dr^n$, hence we obtain
\be\label{evoleqAr}
A'+\frac{A}{2r}\left(1-\frac{1}{f}\right)+4\frac{f n\left(\sqrt{r}n\right)'}{N \sqrt{r}} ={\cal O}V,
\ee
where a prime denotes a derivative with respect to $r$ and the differential operator ${\cal O}$ is
\be
{\cal O}=\frac{r N}{4f}-\frac{\sqrt f N}{2r}\sum_{n=0}^3(-1)^n
\frac{d^n}{dr^n}\left(\frac{r^2}{\sqrt f}\frac{\partial}{\partial f^{(n)}}\right).
\ee
After same algebra, for $f(r)=1-\frac{2 B}{r}$ and $N=1$, we obtain
\bea \label{or}
{\cal O}R&=&\frac{B}{r(r-2B)}\\
{\cal O}R^2&=&0\nn
{\cal O}R^{ij}R_{ij}&=&\frac{B^2}{2r^4(r-2B)}\nn
{\cal O}R^3&=&0\nn
{\cal O}RR^{ij}R_{ij}&=&-\frac{6 B^2(11 B-6 r)}{r^7(r-2B)}\nn
{\cal O}R^i_{~j}R^j_{~k}R^k_{~i}&=&-\frac{3B^2(50 B-27 r)}{2r^7(r-2B)}\nn
{\cal O}R\nabla^{2}R&=&0 \nn
{\cal O}\nabla_i R_{jk} \nabla^i R^{jk}&=& \frac{3B^2(40 B-21 r)}{2r^7(r-2B)}\nonumber.
\eea
Finally, the variation of the action with respect to $N(t)$ gives the so called Hamiltonian constraint
\be
\int_{0}^{\infty}dr \frac{r^2}{\sqrt{f(r)}}(T+V)=0,
\ee
and, using a time redefinition, the lapse function $N(t)$ is set equal to unity ($N(t)=1$).

\section{Analytic solutions} \label{as}

We have to satisfy the following two equations of motion
\be \label{eq1}
f'(r) n(r)=0
\ee
\be \label{eq2}
A'+\frac{A}{2r}\left(1-\frac{1}{f}\right)+\frac{4 f n\left(\sqrt{r}n\right)'}{\sqrt{r}} ={\cal O}V,
\ee
and the Hamiltonian constraint
\be \label{HC}
\int_{0}^{\infty}dr \frac{r^2}{\sqrt{f(r)}}(T+V)=0,
\ee
while the constraint of the spatial curvature ($R=0$) gives
\be
f(r)=1-\frac{2 B}{r}.
\ee
The constraint for $\nu$ (Eq. (\ref{const2})) is verified for $\nu=0$, what we will assume in what follows.

\subsection{Nonzero radial shift function: $n(r)\neq 0$} \label{Sec5.1}

If the shift function does not vanish, we see from the constraint (\ref{eq1}) that $f$ must be constant, and thus $B=0$, such that
\be
f=1.
\ee
As a consequence, $V=0$ (see Eqs. (\ref{or})), and Eq. (\ref{eq2}) can be written as
\be\label{eq3}
r A'+2(rn^2)'=0,
\ee
while Eq. (\ref{HC}) gives the Hamiltonian constraint:
\be\label{HC1}
\int_{0}^{\infty} dr(rn^2)'=0~.
\ee
From Eqs. (\ref{eq3}) and (\ref{HC1}), we obtain then
\be \label{Eq40}
\int_0^\infty dr~rA'(r)=0~.
\ee
The above integral is convergent only if $A(r)$ has the following large and short distance
asymptotic behavior
\begin{itemize}
\item $A(r)\simeq C_{\infty}+C_{A}~ r^{-b}$ for $r\rightarrow +\infty$, with $b>1$;\\
\item $A(r)\simeq C_{0}+ \bar{C}_{A}~ r^{-a}$ for $r\rightarrow 0$, with $a<1$;\\
\end{itemize}
in which $C_{\infty}$, $C_{A}$, $ C_{0}$ and $\bar{C}_{A}$ are arbitrary constants.
An integration by parts, of Eq. (\ref{Eq40}), leads to
\bea\label{intA=0}
\int_0^\infty dr r A'(r)&=&\int_0^\infty dr r \left(A(r)-C_{\infty}\right)'\nonumber \\
&=&\left[r \left(A(r)-C_{\infty}\right)\right]_{0}^\infty-\int_0^\infty dr  \left(A(r)-C_{\infty}\right) \nonumber \\
&=&-\int_0^\infty dr  \left(A(r)-C_{\infty}\right)
\eea
Note, that the fields $A$ and $A+const$ are equivalent in the sense that they give the same $n^2$ from Eq. (\ref{eq3}), hence  without loss of generality we can set
\be \label{Eq42}
C_{\infty}=\lim_{r\rightarrow \infty} A(r)=0
\ee
In this case the Hamiltonian constraint of Eq. (\ref{Eq40}) is equivalent to
\be \label{Eq43}
\int_0^\infty dr  A(r)=0
\ee
In what follows we consider separately two cases 1) $A(r)=0$ which is the minimal choice, 2) $A(r)$ is a function
which satisfy the Hamiltonian constraint of Eq. (\ref{Eq43}) (or the equivalent equation (\ref{Eq40})).

\subsubsection{First case: $A$=0}

When $A=0$, the constraint of Eq. (\ref{Eq43}) is satisfied, hence from Eq. (\ref{eq3}) we obtain
\be
n(r)=\pm\sqrt{\frac{C_{M}}{r}},~~ C_{M}=2 G M c^2
\ee
where $C_{M}$ is a constant of integration. Thus the metric
of Eq. (\ref{metric}) can be written as
\be \label{PG}
ds^2=-c^2 dt^2+\left(dr\pm\sqrt{\frac{2 G M c^2}{r}}dt\right )^2+r^2 (\sin^2\theta d\theta^2+d\phi^2),
\ee
which is the Schwarzschild solution in Painlev\'e-Gullstrand coordinates, see for example Ref. \cite{Tang:2009bu}
and references therein. Note that in this case, the Newtonian potential $\phi(r)$ ($g_{00}=1+2 \phi(r)$)
is proportional to the square of
the radial shift function, according to the equation
\be \label{Newton1}
\phi(r)=-\frac{n^2(r)}{2 c^2}=-\frac{ G M }{r}
\ee
However, we observe that the above expression for the Newtonian potential is not $U(1)$ invariant. In particular it is
the Newton Law for particular Gauge choice $\nu=0$. We can correct this situation if we take into account the coupling between shift functions $N_i$ and matter. In Ref. \cite{daSilva:2010bm} (see also Appendix B)
we see that $N_i$ couple with matter in the $U(1)$ invariant form $N_i-N \nabla_i \nu$, such that
the Newtonian potential should be modified as
\be \label{Newton2}
\phi(r)=-\frac{\left(n(r)-\nabla_r \nu(r)\right)^2}{2 c^2}=-\frac{ G M }{r}.
\ee
The above expression is U(1) invariant, and for the Gauge choice $\nu=0$ we recover Eq. (\ref{Newton1}).

Finally, in the case of the initial HL Gravity model with projectability condition, spherically symmetric solutions have
been studied in Ref. \cite{Tang:2009bu}. The solution we present in this section, for Covariant HL Gravity,
is also a solution of HL Gravity without the $U(1)$ symmetry, but these two models have not the same full spectrum of solutions.
Also, spherically symmetric solutions with nonzero energy momentum tensor (stars), in the case of projectability condition, have been studied in Ref. \cite{Greenwald:2009kp}, and spherically symmetric star solutions are studied in Ref. \cite{Izumi:2009ry}: the main conclusion is that a spherically-symmetric star should include a time-dependent region near the center.

\subsubsection{Second case: $\int_0^{\infty} dr A(r) =0$} \label{Sec5.2.2}

In this situation, the most general solution of Eq. (\ref{eq3}) is
\bea \label{nsqb}
n^2(r)&=&\frac{\tilde C_{M}}{r}-\frac{1}{2} A(r)+\frac{1}{2 r}\int_0^r d\rho~A(\rho)\nn
&=&\frac{\tilde C_{M}}{r}-\frac{1}{2} A(r)-\frac{1}{2 r}\int_r^\infty d\rho~A(\rho),
\eea
where $\tilde C_M$ is a constant,
and we will consider separately the two cases: \textbf{(a)} $\tilde C_M=0$ and \textbf{(b)} $\tilde C_M\ne0$.\\

\nin\textbf{(a)} We have for $\tilde C_M=0$
\be \label{nsqa}
n^2(r)=-\frac{1}{2} A(r)-\frac{1}{2 r}\int_r^\infty d\rho~A(\rho)~.
\ee
Since $A(r)$ behaves as $A(r)\simeq -C_{A} r^{-b}$ for large distances,
Eq. (\ref{nsqa}) gives a modified Newtonian potential when $r\to\infty$
\be\label{phi2ndcase}
\phi(r)=-\frac{n^2(r)}{2 c^2}\simeq -\frac{C_{\phi}}{r^b}, ~~b>1,
\ee
in which
\be
C_{\phi} = \frac{bC_A}{4c^2(b-1)},
\ee
where $C_A>0$ is dimensionful, in order for the potential to have the correct dimensionality, and $C_{\phi}$ is
interpreted as the mass of the compact object.
Also, because $n^2\geq 0$, we have the additional constraint:
\be\label{intAr}
\int_0^r A(\rho)d\rho\geq r A(r),
\ee
and to show a function $A(r)$ satisfying the above constraints can be found, we give here two examples.\\
A first example of function $A(r)$, which satisfies the condition (\ref{intA=0}), is
\be
A_1(r)=-\frac{C_A}{1+r^b}\left( 1-\frac{\gamma_1}{r^{1/b}}\right),~~~~b>1
\ee
where
\be
\gamma_1=\frac{\sin\left( \frac{\pi}{b}-\frac{\pi}{b^2}\right)}{\sin\left( \frac{\pi}{b}\right)}.
\ee
The condition (\ref{intAr}) is then satisfied for all $r$ only if $C_A>0$, such that the potential (\ref{phi2ndcase})
is negative, as expected.
Note that this function is singular for $r=0$, and another example of function $A(r)$ which is
regular at the origin and which satisfies the constraints, is:
\be
A_2(r)= -\frac{C_A}{1+r^{b+1}}\left( r-\gamma_2\right),~~~~b>1
\ee
where
\be
\gamma_2=\frac{\sin\left(\frac{ \pi}{b+1}\right)}{\sin\left(\frac{ 2 \pi}{b+1}\right)}.
\ee
Although we have not check it, by choosing $b$ appropriately closely to unity $b\approx 1$ it may be possible
the above two solutions to pass solar system tests. However, the purpose of this paper is not to set
constraints on the parameter $b$, hence the topic of constructing solutions which can satisfy solar system
tests is left for future investigations. Also, note that the leading order of large distance behavior for the
radial shift function $n(r)$ (or $A(r)$) is fixed by the requirement of recovery of Newton Law in the large
distance limit, but for small distances there is an ambiguity in the exact shape of $n(r)$ (or $A(r)$),
as we see from the above two examples. \\

\nin\textbf{(b)} For $\tilde C_M\ne 0$ we have
\be
n^2(r)=\frac{\tilde C_{M}}{r}-\frac{1}{2} A(r)-\frac{1}{2 r}\int_r^\infty d\rho~A(\rho),
\ee
which corresponds to a qualitatively different situation from the one where $\tilde C_M=0$. Here
the constant $\tilde C_{M}$ is proportional
to the mass of the spherical compact object $\tilde C_M=C_M=2 G M c^2$, and the auxiliary field $A$ determines the
subleading behavior in the asymptotic expansion of $n^2$ for large $r$. As in the previous case $(\tilde C_M=0)$, we
can choose suitably the field $A(r)$ in order to satisfy the Hamiltonian constraint of Eq. (\ref{intA=0}), and
the restriction $n^2>0$.\\
It is explained in \cite{Greenwald:2010fp} that solar system tests
requires a large distance asymptotic behavior for $A$ of the form:
\be
A(r)\simeq \frac{\tilde{C}_{A}}{r^{b}}+\cdots, ~~~~\mbox{when}~~r\rightarrow \infty,
\ee
where the exponent satisfies $b\geq 3$, and the dots represent higher order powers of $1/r$. We find here that the following choice for $A(r)$ has
the required asymptotic behavior, and satisfies the corresponding constraints (the Hamiltonian constraint of
Eq. (\ref{intA=0}) and the requirement $n^2\geq 0$),
\bea \label{A3}
A_3(r)&=&\frac{\tilde{C}_A}{1+r^{b_{1}}}\left( 1-\gamma_3 r^{b_2}\right)~,\\
\mbox{with}&&3\leq b_1, ~~3\leq b_1-b_2, ~~\mbox{and} ~~~-1<b_2\ne0\nonumber
\eea
where $\tilde{C}_A$ is a dimensionful constant and
\be
\gamma_3=\frac{\sin\left( \frac{\pi}{b1}+\frac{\pi b_{2}}{b_{1}}\right)}{\sin\left( \frac{\pi}{b_{1}}\right)}.
\ee
Hence solar system tests do not necessarily impose $A=0$, and the choice for the function $A$ remains an open question.
Furthermore, according to the above examples we see that there is a freedom in the choice of $A$, which correspond to different spherically symmetric solutions, in contrast to standard General Relativity for which spherical symmetry (in the absent of matter) leads to Schwarzschild geometry.

\subsection{Zero radial shift function: $n(r)=0$}\label{Sec5.2}

In this situation, $f(r)=1-2B/r$, with $B\ne0$, and the evolution equation (\ref{evoleqAr}) for $A$ gives
\bea\label{evoleqAu}
\frac{dA}{dr}-\frac{BA}{r(r-2B)}&=&-\frac{c^2B}{r(r-2B)}-\frac{\alpha_2B^2}{2r^4(r-2B)}\\
&&-\frac{6\beta_2B^2}{(r-2B)}\left( \frac{6}{r^6}-\frac{11B}{r^7}\right)\nn
&&-\frac{3\beta_3B^2}{2(r-2B)}\left( \frac{27}{r^6}-\frac{50B}{r^7}\right)\nn
&&+\frac{3\beta_5 B^2}{2(r-2B)}\left( \frac{21}{r^6}-\frac{40B}{r^7}\right) \nonumber.
\eea
The solution of this equation is
\bea \label{A}
A(r)&=&c^2+A_0\sqrt{1-2x}-\frac{\alpha_2}{10B^2}\left(-2+2x+x^2+x^3\right)+\frac{6\beta_2}{B^4}x^6\nn
&&+\frac{\beta_3}{11B^4}\left(-\frac{4}{7}+\frac{4}{7}x+\frac{2}{7}x^2+\frac{2}{7}x^3
+\frac{5}{14}x^4+\frac{x^5}{2}+75x^6\right) \nn
&&-\frac{3 \beta_5}{11B^4}\left(-\frac{4}{7}+\frac{4}{7}x+\frac{2}{7}x^2+\frac{2}{7}x^3
+\frac{5}{14}x^4+\frac{x^5}{2}+20x^6\right),
\eea
where $x=B/r$ and $A_0$ is a constant of integration.
If we set $B=G M$  \footnote{In the case of solution with zero shift function the constant of integration $B$ is proportional to the mass of the compact object.},
the solution of Eq.  (\ref{A}) can be written as an expansion in $x= M/r$ (from now on it is convenient to set $G=1$)
\bea
A(r)&=&c^2+A_0+\frac{\alpha_2}{5M^2}-\frac{4\beta_3}{77M^4}+\frac{12\beta_5}{77M^4} \\
&&+\left(x+\frac{x^2}{2}+\frac{x^3}{2}\right)
\left( -A_0-\frac{\alpha_2}{5M^2}+\frac{4\beta_3}{77M^4}-\frac{12\beta_5}{77M^4}\right) \nn
&&+x^4\left( -\frac{5A_0}{8}+\frac{5\beta_3}{154M^4}-\frac{15\beta_5}{154M^4}\right) \nn
&&+x^5\left( -\frac{7A_0}{8}+\frac{\beta_3}{22M^4}-\frac{3\beta_5}{22M^4}\right) \nn
&&+x^6\left( -\frac{21A_0}{8}+\frac{6\beta_2}{M^4}+\frac{75\beta_3}{11M^4}-\frac{60\beta_5}{11M^4}\right) \nn
&&+{\cal O}(x^7).\nonumber
\eea
in which the constant of integration $A_0$ is chosen to vanish the constant term
\be
c^2+A_0+\frac{\alpha_2}{5M^2}-\frac{4\beta_3}{11M^4}+\frac{12\beta_5}{77M^4}=0,
\ee
Now $A(r)$ can be split to an IR and an UV part according to the equation:
\be
A(r)=A_{IR}(r)+A_{UV}(r)
\ee
where we have set
\bea
A_{IR}(r)&=&c^2\left(1-\sqrt{1-2x}\right)\\
A_{UV}(r)&=&-\left(\frac{\alpha_2}{5M^2}-\frac{4\beta_3}{11M^4}+\frac{12\beta_5}{77M^4}\right)\sqrt{1-2x}\nn
      &&-\frac{\alpha_2}{10M^2}\left(-2+2x+x^2+x^3\right)+\frac{6\beta_2}{M^4}x^6\nn
&&+\frac{\beta_3}{11M^4}\left(-\frac{4}{7}+\frac{4}{7}x+\frac{2}{7}x^2+\frac{2}{7}x^3
+\frac{5}{14}x^4+\frac{x^5}{2}+75x^6\right) \nn
&&-\frac{3 \beta_5}{11M^4}\left(-\frac{4}{7}+\frac{4}{7}x+\frac{2}{7}x^2+\frac{2}{7}x^3
+\frac{5}{14}x^4+\frac{x^5}{2}+20x^6\right),
\eea

\subsubsection{The potential interpretation of $A$} \label{geometric}

According to the original formulation of covariant HL gravity by Horava Melby (HM) in \cite{Horava:2010zj} the field $A$ has a particular
role in the IR limit: more specifically it is promoted as a part of an effective general relativistic metric via the replacement
\begin{equation}
N\rightarrow N-\frac{A_{IR}}{c^2}~.
\end{equation}
Hence the spacetime geometry is determined effectively by the following metric
\begin{equation} \label{metric4}
ds^2=-c^2\left(N^2-\frac{N^{i}N_{i}+2 A_{IR} N}{c^2}\right)dt^2 +2 N^{i}dx_{i}dt+g_{ij}dx^{i}dx^{j},
\end{equation}
which is realized in the nonrelativistic limit $c\rightarrow \infty$, by dropping higher order terms in $1/c^2$, as it is shown initially in \cite{Horava:2010zj} (for details see also Appendix A).

In the above effective metric of Eq. (\ref{metric4}) we have included only the $A_{IR}$ part of $A$ which is consistent with
HM approach. On other hand the $A_{UV}$ part of $A$ can not be included in this metric, as it is subdominant in the $1/c^2$ expansion. Note that an investigation of the physical role of $A_{UV}$ (in HM approach) is beyond the scope in this paper.

It is worth noting that, in the case of solutions with zero shift function ($N_{i}=0$ and $N=1$), the HM approach allows a potential interpretation for the
field $A$. If we take into account that in the IR limit ($r \rightarrow \infty$) $$g_{00}= 1+2\phi+\cdots$$ and compare with Eq. (\ref{metric4})
 above, we obtain that the Newtonian potential $\phi(r)$ is related with $A(r)$ according to the equation
\be
\phi(r)= -\frac{A_{IR}(r)}{c^2}=-\frac{G M}{r}+O\left(\frac{GM}{r}\right)^2~.
\ee
Also the effective metric of Eq. (\ref{metric4}) in the IR limit, in the case of solutions with zero shift function, can be written as
\be \label{soln}
ds^2_{eff}=-c^2\left(1-\frac{2 M}{r}\right) dt^2+\left(1-\frac{2 M}{r}\right)^{-1}dr^2+r^2 d\Omega^2~.
\ee
It is clear that the above metric mimics Schwarzschild geometry in the IR, such that the metric of Eq. (\ref{soln}) passes solar system tests, and no restriction on Horava couplings is necessary (for details see also \cite{Horava:2010zj}). At this point, it should be emphasized that the geometric approach for $A$  is an independent assumption of HM theory, however in this paper we use it mainly to make our solutions (with zero shift function) physically relevant. Finally, the full expression of $A(r)$ for $r\rightarrow\infty$, is given by the expression
\bea \label{ac}
\frac{A(r)}{c^2}&=&\frac{M}{r}+\frac{M^2}{2r^2}+\frac{M^3}{2r^3}\\
&&+\left( \frac{5}{8}+\frac{\alpha_2}{8M^2c^2}\right)\left( \frac{M^4}{r^4}+\frac{7M^5}{5r^5}\right) \nn
&&+\left( \frac{21}{8}+\frac{21\alpha_2}{40M^2 c^2}+\frac{6\beta_2}{M^4 c^2}+\frac{147\beta_3}{22M^4 c^2}-\frac{111\beta_5}{22M^4 c^2}\right)\frac{M^6}{r^6}\nn
&&+{\cal O}(M^7/r^7)~.\nonumber
\eea
We note that the first two corrections, in $M^2/r^2$ and $M^3/r^3$, are independent of the Horava-Lifshitz couplings.
Also, one can see that
corrections of high order derivatives vanish in the nonrelativistic limit ($c\rightarrow \infty$).

\subsubsection{The Hamiltonian constraint} \label{ham}

\begin{figure}[h]
\begin{center}
\includegraphics[width=1 \textwidth, angle=0]{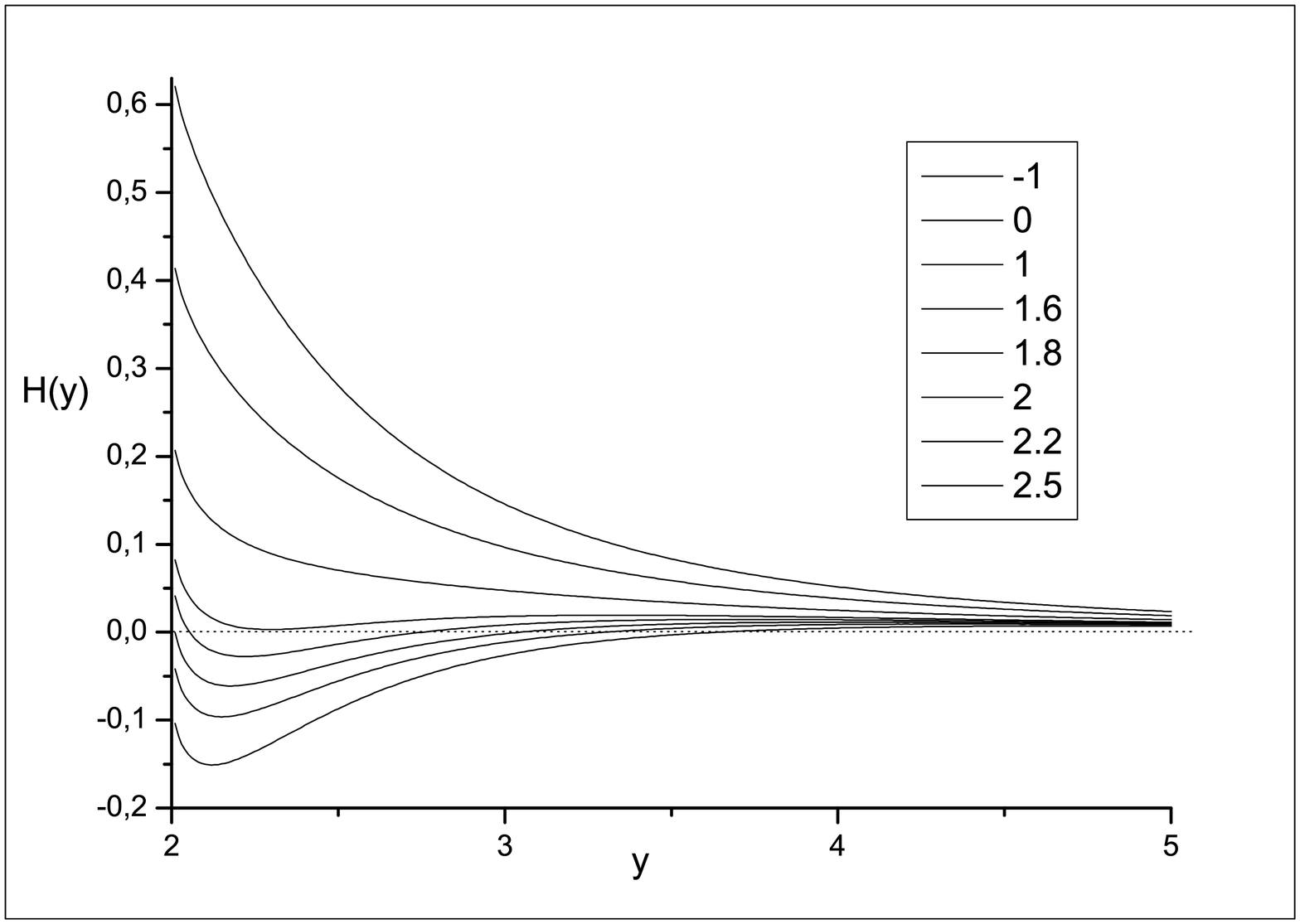}
\end{center}
\caption{\small $H(y)=\alpha_2 C_2(y)- \tilde{\beta}_3 C_3(y)- \tilde{\beta}_5 C_5(y)$ versus $y$, for $\tilde{\beta}_3 =5$ and
increasing values for $\tilde{\beta}_5$, from -1 (upper curve), 0, 1, 1.6, 1.8, 2, 2.2, 2.5 (lower curve).}\label{1}
\end{figure}

In the situation where $n=0$, the Hamiltonian constraint reads
\be\label{condition1}
\int_{0}^\infty dr ~\frac{r^2}{\sqrt{f}} V=0,
\ee
For the case we consider, $f(r)=1-2M/r$, the potential $V$ is
\bea\label{condition2}
V&=&\frac{6\alpha_2M^2}{r^6}-\frac{6\beta_3M^3}{r^9}-\frac{90\beta_5M^2}{r^9}(r-2M),
\eea
In order to keep the integrand function in Eq. (\ref{condition1}) real, we have to introduce
a spherical gap in space,
centered on the black hole and including the horizon, assuming that the radial coordinate $r$ has the minimum value $L$,
where $L\geq 2 M$. An equivalent alternative
\footnote{The authors would like to thank Alex Kehagias for this suggestion.}
is for example to introduce the new coordinate $u\geq 0$, defined
by $r=\sqrt{u^2+L^2}$, and to express the whole problem in terms of $u$ instead of $r$.
Note that space in this situation is still simply connected, such that this non-vanishing length $L$
does not introduce topological defects. In this case we have
\bea\label{int1}
\int_{L}^\infty dr~ \frac{r^2}{\sqrt{f}} V
&=&6M^2\int_{L}^\infty \frac{dr}{\sqrt{r-2M}}\left( \frac{\alpha_2}{r^{7/2}}-\frac{\beta_3M}{r^{13/2}}
+\frac{15\beta_5}{r^{13/2}}(r-2M)\right)\nn
&=&M^{-1}\alpha_2 C_2- M^{-3}\beta_3 C_3-M^{-3}\beta_5 C_5,
\eea
where the dimensionless functions $C_2,C_3,C_5$ are given by the equations
\bea \label{const1}
C_2(y)&=&\frac{4}{5}-\frac{2 \sqrt{y-2}}{5 y^{5/2}}(3+2y+2y^2)\\
C_3(y)&=&\frac{16}{231}-\frac{2\sqrt{y-2}}{231 y^{11/2}}(63+35 y +20y^2 \nn
&&~~~~~~~~~~~~~~~~~~~~~~~~~+12 y^3 +8y^4+8 y^5) \nn
C_5(y)&=&\frac{16}{77 } -\frac{2(y-2)^{3/2}}{77 y^{11/2}}(315+140 y +60y^2 \nn
&&~~~~~~~~~~~~~~~~~~~~~~~~~+24 y^3+8 y^4), \nonumber
\eea
where we have set $y=L/M$.
The Hamiltonian constraint leads to the following algebraic equation
\bea \label{leq}
 C_2(y)- \tilde{\beta}_3 C_3(y)- \tilde{\beta}_5 C_5(y)=0,
\eea
in which we have set $\tilde{\beta}_3=\beta_3 \alpha_2^{-1} M^{-2}$ and $\tilde{\beta}_5=15 \beta_5 \alpha_2^{-1} M^{-2}$. We can determine the lower limit $L$ by solving numerically Eq. (\ref{leq}), for certain values of $\tilde{\beta}_3$ and $\tilde{\beta}_5$. Note, that a detailed investigation of Eq. (\ref{leq}) is rather involved and unnecessary for this study. However, we have performed computations for specific values of $\tilde{\beta}_3$ and $\tilde{\beta}_5$, and we present our results in Fig. \ref{1}. In this figure he have plotted the function $$H(y)= C_2(y)- \tilde{\beta}_3 C_3(y)- \tilde{\beta}_5 C_5(y)$$ versus $y$, for fixed $\tilde{\beta}_3 =5$, and several values of $\tilde{\beta}_5 =-1,0,1,1.6,1.8,2,2.2,2.5$. We observe that for $\tilde{\beta}_5 \leq 1.6$ Eq. (\ref{leq}) has no real solutions, for $1.6 <\tilde{\beta}_5\leq 1.8$  Eq. (\ref{leq}) has two real solutions, and for $1.8<\tilde{\beta}_5$ Eq. (\ref{leq}) has one real solution. This situation seems to be quite general, as we have performed computations for other values of $\tilde{\beta}_3$, which are not presented in this paper, and we observed the same behavior for the function $H(y)$. According to the above mentioned results, there is a region of the parameter space for which there is no lower limit $L$, which vanish the Hamiltonian constraint of Eq. (\ref{leq}). Also, there is a region in which there is an arbitrariness is the choice of $L$, as Eq. (\ref{leq}) has two distinct solutions for $y=L/M$. Finally, there is a region for which $L$ is determined uniquely by the couplings $\tilde{\beta}_3$, and $\tilde{\beta}_5$ of the spherical symmetric object, as Eq. (\ref{leq}) has one real solution.

It would be nice if we could determine the range of $r$ by physical and geometrical
considerations. However, the spherically symmetric solutions with zero shift function should satisfy the Hamiltonian constraint. 
Note that the potential
in Eq. (\ref{condition2}) diverges for $r=0$, so that it is unavoidable to introduce a lower limit $L$ in the corresponding 
integral of Eq. (\ref{int1}).
As we show explicitly above, this lower limit is determined by the free parameters of the model, if we try to satisfy the 
Hamiltonian
constraint of Eq. (\ref{int1}) (or of Eq. (\ref{condition1})). It seems that the range of acceptable values for $r$ can not 
be determined geometrically,
for spherically symmetric solutions with zero shift function in Covariant Horava Lifshitz
Gravity, unlike the case of standard black hole solutions in General Relativity. However, we think that this topic needs further investigation.

\section{Conclusions} \label{con}

We studied the most general case of spherically symmetric vacuum solutions in the framework
of Covariant Horava Lifshitz Gravity. More specifically, we found two classes of spherically symmetric
solutions: {\it (i)} with nonzero radial shift function (see Sec. \ref{Sec5.1}) and
{\it (ii)} with zero radial shift function (see Sec. \ref{Sec5.2}). We show that solutions of class {\it (ii)} becomes
physically relevant if we adopt the position of HM for the geometrical role of $A$ in the IR, while solutions
of class {\it (i)} are consistent with observations only if $A$ is independent of spacetime geometry.

In the case of solutions with nonzero radial shift function we would like to mention the freedom
of the non-dynamical field $A(r)$, although the latter has to satisfy constraint equations, as we see in
Sec. \ref{Sec5.1}. In particular, $A(r)$ can be a constant, or it must satisfy a set of constraints
which are presented in Sec. \ref{Sec5.2.2}. This class of solutions permits an alternative interpretation
for the field $A$ other than that of HM, as the Newtonian potential for large distances is recovered by the radial shift function. In the latter case - situation {\it (ii)} - one needs to interpret geometrically the $U(1)$ transformation as a space time symmetry, in order to recover the Newtonian potential in the non-relativistic limit, which in the situation {\it(i)} is not necessary.\\
Note that, in order to recover the standard Schwarzschild geometry, expressed in Painlev\'e-Gullstrand coordinates,
one needs to take the solution for which $A$ is constant. For the other solutions, where $A(r)$ is not constant and satisfies the
constraints in Sec.\ref{Sec5.2.2}, we obtain solutions which obey a modified power law for the Newtonian potential,
with a leading term $\phi(r)\sim r^{-b}$ and $b>1$. These solutions may possibly be consistent with solar system tests if $b$ is chosen appropriately
closely to unity ($b\approx 1$), but this would still need to be checked. However, in Sec.\ref{Sec5.2.2} we present a
solution, which can satisfy solar system tests and corresponds to a nonconstant field $A$ (see Eq. (\ref{A3})), and a
spacetime geometry which agree with Schwarzschild metric only asymptotically.\\
Note that, although the leading order of large distance behavior for the radial shift function of $A(r)$ is fixed
by the requirement of recovery of Newton Law in the large distance limit, for small distances there is an
ambiguity in the exact shape of $A(r)$, as we see from the examples in Sec. \ref{Sec5.2.2}. \\Finally, for solutions with non-zero radial component of shift functions, higher order spatial curvature corrections in the action do not play a role, and the coupling constants in Eq. (\ref{potential}) do not appear in this class of solutions.

In the case of zero radial shift function, in contrast with the previous case,
higher order curvature terms do not vanish and, as a result, the corresponding higher order couplings appear in the solution for
the auxiliary field $A$. In Sec. \ref{geometric} we explain why only the IR part of $A$ possesses a geometric interpretation in the IR, while the role of the UV part of $A$, in the framework of HM view, remains ambiguous.
\\In addition, this situation is characterized by a difficulty related to the Hamiltonian constraint.
In order to satisfy the latter, we had to assume the existence of a spherical gap in space (with radius $L$), centered on the spherical compact object. The radius $L$ of the gap can be determined by an algebraic equation which can be solved only numerically, as it is shown in Sec. \ref{ham}. Note, that there are regions of the free parameter space for which this equation has no solutions, or has more than one solution (in particular two). In this sense, for the case of two solutions, there is an arbitrariness in the choice of $L$. In addition, in this class the auxiliary field $A$ is determined unambiguously by the corresponding constraint (for details see Secs. \ref{sps} and \ref{Sec5.2}).
Also, note that this class of solutions can pass solar system tests, failing to impose restrictions to Horava couplings.\\
The main conclusion of this paper, is that beyond the geometrical approach of HM, there is an alternative view for the field $A$ which is also consistent with observations. In the latter case the field $A$ is considered to be independent of the spacetime geometry, in contrast to HM theory, and the Newton law is reproduced by the nonzero shift function of the solutions of class {\it (i)}. Finally, note that the field $A$ is not completely fixed for solutions of class {\it (i)}, so a study on this topic may be a topic for future investigation.

\vspace{0.5cm}

\nin{\bf Acknowledgments} We would like to thank K. Farakos and A. Kehagias
for numerous valuable discussions. This work is supported by the Royal Society, UK.

\vspace{0.5cm}

\nin{\bf Note added in proof} We would like to acknowledge that the alternative view for the field 
$A$ has been also proposed independently in the parallel paper \cite{Greenwald:2010fp}, 
which appeared on the arXiv almost simultaneously with this work.

\section*{Appendix A: Geometrical interpretation of the $U(1)$ symmetry} \label{appA}

We consider the standard metric in the ADM form
\begin{equation} \label{metric1}
ds^2=-c^2\left(N(t)^2-\frac{N^{i}N_{i}}{c^2}\right)dt^2 +2 N^{i}dx_{i}dt+g_{ij}dx^{i}dx^{j},
\end{equation}
in which the lapse function $N(t)$ is assumed to be only a function of time.
We can promote the auxiliary field $A(x,t)$ as a part of spacetime geometry, and simultaneously introduce a new spacetime dependent
the lapse function, by performing the replacement
$$
N(t)\rightarrow N(t)-\frac{A(x,t)}{c^2}.
$$
The above metric can be written as
\begin{equation} \label{metric2}
ds^2=-c^2\left(N^2-\frac{N^{i}N_{i}+2 AN}{c^2}\right)dt^2 +2 N^{i}dx_{i}dt+g_{ij}dx^{i}dx^{j},
\end{equation}
where we have dropped the higher order term $A^2/c^4$, as we are interested for the nonrelativistic limit $c\rightarrow \infty$.

It is easy to see, in the limit where $c\rightarrow \infty$,
that the metric (\ref{metric2}) is invariant under a spacetime dependent reparametrization of time, according to
\begin{equation}
t'=t+\frac{\varepsilon(x,t)}{c^2 },~~~~~~~~~~ x'=x,
\end{equation}
if this transformation is accompanied by the $U(1)$ symmetry
\begin{eqnarray}\label{U1}
N'_{i}(x,t)&=&N_{i}(x,t)+N^2\nabla_{i}\varepsilon \nonumber \\ 
A'(x,t)&=&A(x,t)+\dot{\varepsilon}N+\varepsilon\dot{N}
-NN^{i}\nabla_{i}\varepsilon,
\end{eqnarray}
where
$$
\varepsilon(x,t)=\frac{\alpha(x,t)}{N}.
$$
(Note that, under this $U(1)$ transformation, the fields $N(t)$ and $g_{ij}(x,t)$ remain unaltered).
Indeed, if we consider the general relativistic metric in the $t'$ coordinate
\begin{equation}\label{metric3}
ds^2=-c^2\left(N^2-\frac{N'^{i}N'_{i}+2 A'N}{c^2}\right)(dt')^2 +2 N'^{i}dx_{i}dt'+g_{ij}dx^{i}dx^{j},
\end{equation}
where all the fields depend on $t'$, the $U(1)$ transformation of Eqs. (\ref{U1}), together with
\begin{equation}
N(t')=N\left(t+\frac{\varepsilon(x,t)}{c^2 }\right)\simeq N(t)+\frac{\dot{N} \varepsilon(x,t)}{c^2 },
\end{equation}
\begin{equation}
dt'=\left(1+\frac{\dot{\varepsilon}}{c^2}\right) dt+\frac{\nabla_{i}\varepsilon dx^i}{c^2}+O\left(\frac{1}{c^4}\right),
\end{equation}
show that the metrics (\ref{metric3}) and (\ref{metric2}) are equivalent, up to
higher order terms in $1/c$.

\section*{Appendix B: $U(1)$ symmetry by a minimal substitution} \label{appB}

In Ref. \cite{daSilva:2010bm} is proposed a minimal substitution mechanism, which can be used to extend the
Gauge symmetry of any $Diff({\cal M},{\cal F })$ invariant action. This mechanism is bases on the observation that the following
quantities are invariant under the $U(1)$ transformation,
\be
\delta_{a}\left(N_{i}-N \nabla_{i}\nu\right)=0,~~~\delta_{a}(A-a)=0
\ee
where $a$ is defined as
\be
a=\dot{\nu}-N^{j}\nabla\nu_{j}+\frac{N}{2}\nabla_{j}\nu\nabla^{j}\nu~.
\ee
We consider, a $Diff({\cal M},{\cal F})$ invariant action of the form:
\be
S[N,N_{i},g_{ij},\psi_{n}]=S_{HL}[N,N_{i},g_{ij}]+S_{m}[N,N_{i},g_{ij},\psi_{n}],
\ee
in which the first term $S_{HL}$ is the standard HL action
\be \label{HLA}
S_{HL}[N,N_{i},g_{ij}]=\frac{2}{\kappa^2}\int dt d^3x\sqrt{g}N\left[ K_{ij}K^{ij}-\lambda K^2-V\right],
\ee
and the second term $S_m$ represents the interaction between the external fields $\psi_{n}$ and the gravitational fields  $N,~N_{i},~g_{ij}$.
The fields $\psi_{n}$ may be, for example, scalar or vector fields.
The action $S[N,N_{i},g_{ij},\psi_{n}]$ can be promoted to a manifestly $U(1)$
invariant action $\hat{S}[N,N_{i},g_{ij},\psi_{n},A,\nu]$, if we perform the replacement $N_{i}\rightarrow N_{i}-N \nabla_{i}\nu$,
and simultaneously add an extra term which depends only on $(A-a)$, according to the equation:
\be
\hat{S}[N,N_{i},g_{ij},\psi_{n},A,\nu]=S[N,N_{i}-N \nabla_{i}\nu,g_{ij},\psi_{n}]+\int dtd^3x \sqrt{g} Z(\psi_{n},g_{ij})(A-a)
\ee
where $Z(\psi_{n},g_{ij})$, with dimension $[Z]=2$, is the most general operator which is invariant under $Diff({\cal M},{\cal F})$ and
respects renormalizability requirements. This minimal substitution mechanism give naturally
the answer of how one can couple matter with the auxiliary fields $A$ and $\nu$.

As shown in Ref. \cite{daSilva:2010bm}, in the absent of the external field $\psi_n$, if we set $\lambda=1$, the manifestly
$U(1)$ invariant action  $\hat{S}[N,N_{i},g_{ij},0,A,\nu]$ is identical with the action of Covariant HL Gravity, as it is given by
Eq. (\ref{action}) above, or equivalently we can write:
\bea \label{hala}
\hat{S}&=&S_{HL}[N,N_{i}-N \nabla_{i}\nu,g_{ij}]-\frac{2}{\kappa^2}\int dtd^3x \sqrt{g} (A-a)(R-\Omega) \nonumber\\
&=&\frac{2}{\kappa^2}\int dt d^3x\sqrt{g}\Big\lbrace N\left[ K_{ij}K^{ij}-K^2-V+\nu\Theta^{ij}(2K_{ij}+\nabla_i\nabla_j\nu)\right]\nn
&&~~~~~~~~~~~~~~~~~~~~~~~~~~~~~~~~~~~~~~~~~~~~~~~~-A(R-2\Omega)\Big\rbrace ,
\eea
Note that, if we set $\psi_{n}=0$, we obtain
$$
Z(0,g_{ij})=-\frac{2}{\kappa^2}(R-\Omega).
$$
Finally, we observe that this derivation of the Covariant HL action does not require $\lambda=1$, which implies that
$U(1)$ symmetry can not force the value of $\lambda$ to be $\lambda=1$. However, for $\lambda\neq 1$, the action of Eq. (\ref{hala}) above is
modified by a term of the form:
\be
S_{\lambda}=\frac{2}{\kappa^2}\int dt d^dx \sqrt{g} N(1-\lambda) (K+\triangle\nu)^2~.
\ee


\begin{thebibliography}{99}

 \bibitem{Horava:2008ih}
  P.~Horava,
  JHEP {\bf 0903} (2009) 020
  [arXiv:0812.4287 [hep-th]];
  P.~Horava,
  Phys.\ Rev.\  D {\bf 79}, 084008 (2009)
  [arXiv:0901.3775 [hep-th]].

\bibitem{Kiritsis:2009sh}
  E.~Kiritsis and G.~Kofinas,
  Nucl.\ Phys.\  B {\bf 821} (2009) 467
  [arXiv:0904.1334 [hep-th]].

\bibitem{Calcagni:2009ar}
  G.~Calcagni,
  JHEP {\bf 0909}, 112 (2009)
  [arXiv:0904.0829 [hep-th]].

\bibitem{Tang:2009bu}
  J.~Z.~Tang and B.~Chen,
  Phys.\ Rev.\  D {\bf 81}, 043515 (2010)
  [arXiv:0909.4127 [hep-th]].

\bibitem{Lu:2009em}
  H.~Lu, J.~Mei and C.~N.~Pope,
  Phys.\ Rev.\ Lett.\  {\bf 103}, 091301 (2009)
  [arXiv:0904.1595 [hep-th]].

\bibitem{Kehagias:2009is}
  A.~Kehagias and K.~Sfetsos,
  Phys.\ Lett.\  B {\bf 678} (2009) 123
  [arXiv:0905.0477 [hep-th]];
  C.~Germani, A.~Kehagias and K.~Sfetsos,
  JHEP {\bf 0909}, 060 (2009),
arXiv:0906.1201 [hep-th].

\bibitem{Cai:2009pe}
  R.~G.~Cai, L.~M.~Cao and N.~Ohta,
  Phys.\ Rev.\  D {\bf 80} (2009) 024003
  [arXiv:0904.3670 [hep-th]].

\bibitem{Kiritsis:2009rx}
  E.~Kiritsis and G.~Kofinas,
  JHEP {\bf 1001} (2010) 122
  [arXiv:0910.5487 [hep-th]].

\bibitem{Kiritsis:2009vz}
  E.~Kiritsis,
  Phys.\ Rev.\  D {\bf 81} (2010) 044009
  [arXiv:0911.3164 [hep-th]].

\bibitem{Koutsoumbas:2010pt}
  G.~Koutsoumbas, E.~Papantonopoulos, P.~Pasipoularides and M.~Tsoukalas,
  arXiv:1004.2289 [hep-th].

\bibitem{Koutsoumbas:2010yw}
  G.~Koutsoumbas and P.~Pasipoularides,
  Phys.\ Rev.\  D {\bf 82} (2010) 044046
  [arXiv:1006.3199 [hep-th]].

\bibitem{Cai:2009in}
    Y.~F.~Cai and E.~N.~Saridakis,
  JCAP {\bf 0910}, 020 (2009)
  [arXiv:0906.1789 [hep-th]];
  S.~Dutta and E.~N.~Saridakis,
  JCAP {\bf 1001} (2010) 013
  [arXiv:0911.1435 [hep-th]];
  G.~Leon and E.~N.~Saridakis,
  JCAP {\bf 0911} (2009) 006
  [arXiv:0909.3571 [hep-th]].


\bibitem{Capasso:2009ks}
  D.~Capasso and A.~P.~Polychronakos,
  Phys.\ Rev.\  D {\bf 81} (2010) 084009
  [arXiv:0911.1535 [hep-th]].

\bibitem{Ghodsi:2009zi}
  A.~Ghodsi and E.~Hatefi,
  Phys.\ Rev.\  D {\bf 81}, 044016 (2010)
  [arXiv:0906.1237 [hep-th]].

\bibitem{Greenwald:2009kp}
  J.~Greenwald, A.~Papazoglou and A.~Wang,
  Phys.\ Rev.\  D {\bf 81} (2010) 084046
  [arXiv:0912.0011 [hep-th]].


\bibitem{Greenwald:2010fp}
  J.~Greenwald, V.~H.~Satheeshkumar and A.~Wang,
  arXiv:1010.3794 [hep-th].

\bibitem{Izumi:2009ry}
  K.~Izumi and S.~Mukohyama,
  Phys.\ Rev.\  D {\bf 81} (2010) 044008
  [arXiv:0911.1814 [hep-th]].

\bibitem{Majhi:2009xh}
  B.~R.~Majhi,
  Phys.\ Lett.\  B {\bf 686} (2010) 49
  [arXiv:0911.3239 [hep-th]].

\bibitem{Myung:2009dc}
  Y.~S.~Myung and Y.~W.~Kim,
  arXiv:0905.0179 [hep-th].

 \bibitem{Cai:2009}
 R.~G.~Cai, L.~M.~Cao and N.~Ohta,
 Phys.\ Lett.\  B {\bf 679} (2009) 504
 [arXiv:0905.0751 [hep-th]];
R.~G.~Cai and N.~Ohta,
 Phys.\ Rev.\  D {\bf 81} (2010) 084061
 [arXiv:0910.2307 [hep-th]],

\bibitem{Konoplya:2009ig}
 R.~A.~Konoplya,
 Phys.\ Lett.\  B {\bf 679}, 499 (2009)
 [arXiv:0905.1523 [hep-th]].

\bibitem{Lee:2009rm}
  H.~W.~Lee, Y.~W.~Kim and Y.~S.~Myung,
  arXiv:0907.3568 [hep-th].

\bibitem{Gruss:2010zc}
  E.~Gruss,
  arXiv:1005.1353 [hep-th].


\bibitem{Takahashi:2009wc}
 T.~Takahashi and J.~Soda,
 Phys.\ Rev.\ Lett.\  {\bf 102}, 231301 (2009)
 [arXiv:0904.0554 [hep-th]].

\bibitem{Park:2009zra}
  M.~i.~Park,
  JHEP {\bf 0909} (2009) 123
  [arXiv:0905.4480 [hep-th]].

\bibitem{Colgain:2009fe}
  E.~O.~Colgain and H.~Yavartanoo,
  JHEP {\bf 0908} (2009) 021
  [arXiv:0904.4357 [hep-th]].


\bibitem{Iorio:2009qx}
  L.~Iorio and M.~L.~Ruggiero,
  arXiv:0909.2562 [gr-qc].



\bibitem{Horava:2008jf}
 P.~Horava,
 arXiv:0811.2217 [hep-th].

\bibitem{Visser:2009fg}
  M.~Visser,
  Phys.\ Rev.\  D {\bf 80} (2009) 025011
  [arXiv:0902.0590 [hep-th]];
  M.~Visser,
  arXiv:0912.4757 [hep-th].


\bibitem{Alexandre:2009sy}
  J.~Alexandre, K.~Farakos, P.~Pasipoularides and A.~Tsapalis,
  Phys.\ Rev.\  D {\bf 81} (2010) 045002
  [arXiv:0909.3719 [hep-th]];
J.~Alexandre, N.~E.~Mavromatos and D.~Yawitch,
  arXiv:1009.4811 [hep-ph].



\bibitem{Alexandre:2010hp}
  J.~Alexandre, K.~Farakos and A.~Tsapalis,
  Phys.\ Rev.\  D {\bf 81} (2010) 105029
  [arXiv:1004.4201 [hep-th]].

\bibitem{Anagnostopoulos:2010gw}
  K.~Anagnostopoulos, K.~Farakos, P.~Pasipoularides and A.~Tsapalis,
  arXiv:1007.0355 [hep-th].


\bibitem{Orlando:2009az}
  D.~Orlando and S.~Reffert,
  Phys.\ Lett.\  B {\bf 683} (2010) 62
  [arXiv:0908.4429 [hep-th]].




\bibitem{Charmousis:2009tc}
  C.~Charmousis, G.~Niz, A.~Padilla and P.~M.~Saffin,
  JHEP {\bf 0908} (2009) 070
  [arXiv:0905.2579 [hep-th]].

\bibitem{Li:2009bg}
  M.~Li and Y.~Pang,
  JHEP {\bf 0908}, 015 (2009)
  [arXiv:0905.2751 [hep-th]].


\bibitem{Blas:2009qj}
  D.~Blas, O.~Pujolas and S.~Sibiryakov,
  arXiv:0909.3525 [hep-th].


\bibitem{Papazoglou:2009fj}
  A.~Papazoglou and T.~P.~Sotiriou,
  Phys.\ Lett.\  B {\bf 685} (2010) 197
  [arXiv:0911.1299 [hep-th]].

\bibitem{Kimpton:2010xi}
  I.~Kimpton and A.~Padilla,
  arXiv:1003.5666 [hep-th].

\bibitem{Bellorin:2010je}
  J.~Bellorin and A.~Restuccia,
  arXiv:1004.0055 [hep-th].

\bibitem{Pons:2010ke}
  J.~M.~Pons and P.~Talavera,
  arXiv:1003.3811 [gr-qc].

\bibitem{Horava:2010zj}
  P.~Horava and C.~M.~Melby-Thompson,
  Phys.\ Rev.\  D {\bf 82} (2010) 064027
  [arXiv:1007.2410 [hep-th]].



\bibitem{daSilva:2010bm}
  A.~M.~da Silva,
  arXiv:1009.4885 [hep-th].



\bibitem{Sotiriou:2009bx}
  T.~P.~Sotiriou, M.~Visser and S.~Weinfurtner,
  JHEP {\bf 0910}, 033 (2009),
arXiv:0905.2798 [hep-th];
  T.~P.~Sotiriou, M.~Visser and S.~Weinfurtner,
  Phys.\ Rev.\ Lett.\  {\bf 102} (2009) 251601
  [arXiv:0904.4464 [hep-th]].

\bibitem{Wang:2010wi}
  A.~Wang and Y.~Wu,
  arXiv:1009.2089 [hep-th].




\end{thebibliography}
\end{document}